\documentclass[preprint,12pt]{elsarticle}

\usepackage{amssymb, pgf-pie}
\usepackage{float}
\usepackage{amsmath}
\usepackage{multirow}
\usepackage{subcaption}
\usepackage{url}
\usepackage{hyperref}
\usepackage{natbib}

\journal{International Journal of Human-Computer Studies}

\begin{document}

\begin{frontmatter}



\title{Generation Z’s Ability to Discriminate Between AI-generated and Human-Authored Text on Discord}


\author[inst1]{Dhruv Ramu}

\affiliation[inst1]{,
            addressline={Neev Academy}, 
            city={Bengaluru},
            postcode={560037}, 
            state={Karnataka},
            country={India}}
\affiliation[inst2]{
            addressline={Harvard University}, 
            city={Cambridge},
            postcode={02138}, 
            state={Massachusetts},
            country={United States}}
\affiliation[inst3]{
            addressline={Harvard Medical School}, 
            city={Boston},
            postcode={02115}, 
            state={Massachusetts},
            country={United States}}

\author[inst2]{Rishab Jain}
\author[inst3]{Aditya Jain}

\begin{abstract}
The growing popularity of generative artificial intelligence (AI) chatbots such as ChatGPT is having transformative effects on social media. As the prevalence of AI-generated content grows, concerns have been raised regarding privacy and misinformation online. Among social media platforms, Discord enables AI integrations---making their primarily “Generation Z” userbase particularly exposed to AI-generated content. We surveyed Generation Z aged individuals (n $=$ 335) to evaluate their proficiency in discriminating between AI-generated and human-authored text on Discord. The investigation employed one-shot prompting of ChatGPT, disguised as a text message received on the Discord.com platform. We explore the influence of demographic factors on ability, as well as participants’ familiarity with Discord and artificial intelligence technologies. We find that Generation Z individuals are unable to discern between AI and human-authored text (p $=$ 0.011), and that those with lower self-reported familiarity with Discord demonstrated an improved ability in identifying human-authored compared to those with self-reported experience with AI (p $<<$ 0.0001). Our results suggest that there is a nuanced relationship between AI technology and popular modes of communication for Generation Z, contributing valuable insights into human-computer interactions, digital communication, and artificial intelligence literacy.
\end{abstract}




\end{frontmatter}


\section{Introduction}
\label{sec:introduction}

Artificial intelligence (AI) technologies have permeated various facets of society, leading to transformative changes in communication platforms and digital interactions~\cite{Makridakis_2017}. As AI systems become increasingly sophisticated, they are capable of generating human-like media, blurring the lines between AI-generated and human-authored text.  Social media (which includes Voice over IP (VoIP) and instant messaging) has become the most ubiquitous form of communication in the world with AI-generated text running rampant due to anonymity and automation~\cite{Gupta_Varol_Zhou_2023}. AI-generated text on social media can be used to propagate false information, manipulate public opinion, and even impersonate real people~\cite{bian2023drop}.

Generation Z, comprising individuals born between 1997 and 2012~\cite{dimock2019defining}, emerges as the generation most vulnerable to the proliferation of AI-generated media and its associated technologies~\cite{Vinichenko-viewsofgenz}. Their exposure to digital technologies and high adoption of social media expose them to a myriad of AI-based media, ranging from deepfake videos to personalized algorithmic news~\cite{whyte2020deepfake}. As one of the youngest generations, Generation Z is uniquely positioned to experience and adapt to the proliferation of AI-generated content within the framework of their educational and professional pursuits. In this context, it is critical to assess how Generation Z perceives AI-generated and human-authored text on communication platforms. 

Discord is a widely popular social media platform among Generation Z, with 36.10\% of Discord’s user base being from this age range. This is approximately 203.2 million Generation Z users, and it outpaces other platforms in terms of monthly messages, active users, and user acquisition with a steadily increasing growth rate of  23.46\%~\cite{Ceci_2023}. Discord has emerged as a favored platform for this demographic due to its versatile features that cater to various interest-based communities, deviating from its original focus of `gaming'~\cite{9575079}. Discord is unique in its offerings with the ability to send multimedia messages, voice chat using VoIP, and video call. Discord also facilitates users to create their own persistent chat channels, sizing from a few people to hundreds of thousands of members~\cite{moffitt2021discord}. Through the Discord application programming interface (API), bots can be created to interact with users~\cite{abhinand2022study}. These bots can send and receive messages in a channel, like a human user, and have increasingly become powered by large language models such as GPT-3~\cite{dale_2021}. Popular bots such as `MEE6' and Discord's own `Clyde' have already integrated OpenAI's text and image generation features~\cite{clyde} .

It has been shown that widely-used generative AI chatbots built from large language models (LLMs) such as ChatGPT have limitations and unintended biases. ChatGPT may produce text containing bias, including cultural, language, clickbait, confirmation, source, and novelty biases, that emerge from data during the training of the model \cite{jain2023generative,ray2023chatgpt}. Yet, LLMs can produce coherent and contextually relevant text, making it challenging for users to distinguish between AI-generated and human-authored messages. LLMs also include outdated knowledge due to their knowledge cutoff, poor ability to discern factual accuracy, limited emotional intelligence, and inconsistency in quality \cite{jain2023generative,kocon2023chatgpt}. If outputs from LLMs such as ChatGPT are to be used among accounts on social media, it is important that there is awareness of these limitations. ChatGPT has been shown to prioritize content that aligns with beliefs it is exposed to during training, which can be polarizing and reinforce an ideological bias towards dominant opinions found in its training data~\cite{jain2023generative,ray2023chatgpt}. LLMs can allow bad actors to synthetically generate text in ways that mimic the style and substance of human-created news stories~\cite{kreps2022all}. AI-generated content can be inherently deceptive due to advancements in natural language processing algorithms. This raises concerns regarding the potential consequences of misidentifying the source of a message, particularly in sensitive or critical contexts, where trust and credibility are paramount~\cite{denny2023can}.

Discriminating between AI-generated and human-authored content is difficult. Automated techniques for AI text detection face significant challenges and can not accurately and consistently detect AI-generated content~\cite{pegoraro2023chatgpt}. Further, jailbreaks and redteaming efforts have been employed to reliably fool automated detection software. OpenAI Text Classifier, GPTZero, and Writer.com’s AI text detector, are unable to detect AI-generated text after being translated with Google Translate and have high false positive rates~\cite{chaka2023detecting}. Confidence in these tools is low, with OpenAI pulling their own tool from the public due to its 26\% true positive rate when classifying AI-generated text~\cite{openaiclassifier}. Paraphrasing tools, where a paraphraser is applied on top of the output from an LLM, also reliably bypass automated detection by altering the semantics and linguistic structure of the original text, making it difficult for traditional detection techniques to recognize the underlying patterns and signatures used for AI identification. Bad-faith actors can use ChatGPT to replicate particular writing styles, use clever prompt design to increase perplexity, and make minute manual changes to texts that break a range of detectors~\cite{sadasivan2023aigenerated}. 

This study employs an experimental approach, assessing participants’ (n $=$ 335) accuracy in discriminating between AI-generated or human-authored text. By examining the influence of factors such as education, familiarity with AI technology, and experience with Discord, the study aims to provide insight into potential differences in members of Generation Z's perception of AI-generated text on social media. The findings are expected to contribute to the broader discourse on AI-driven communication and offer valuable insights into the implications of LLMs on the authenticity and trustworthiness of human-computer interactions for Generation Z.

\section{Related Works}
\label{sec:relatedworks}
Prior research has examined the trustworthiness of AI systems when implemented in hiring processes, college admissions, and mortgage approval~\cite{gonzalez2022allying,10.1145/3290605.3300469}. However in many of these studies, only when the participants were exposed to both AI and human-generated texts, and informed about the possibility of AI-generated content, did they suspect it (an example of the ‘Replicant Effect’)~\cite{10.1145/3290605.3300469}. It has also been shown that users distrust content which they believe to be AI generated~\cite{10.1145/3334480.3382842}, but when ``anthropomorphic attributes'' are applied, there is perceived warmth which positively contributes to trust~\cite{cheng2022human,folstad2018makes}.


There is limited research observing whether humans can discriminate between human and AI generated content. One study showed that participants fail to distinguish between AI-generated poetry and the works of Maya Angelou---however, this was using a less advanced GPT-2 model which is very limited compared to GPT-3 or GPT-3.5 \cite{kobis2021artificial}. These participants were more averse to the AI-generated work regardless of whether they knew whether it was generated by AI or not. Other writing, namely scientific abstracts, were analyzed using ‘GPT2-Output-Detector’ and blinded human reviewers \cite{gao2023comparing}. 68\% of human reviewers correctly identified AI-generated text, whilst 14\% falsely found original abstracts to be AI-generated. It is also important to note that these abstracts, when AI generated, had far more potential to have factual inaccuracies, ambiguity, and apparent monotony or low perplexity due to length~\cite{gao2023comparing}. However, similar trends would not be observed when the text to compare is shorter and the reviewer is not an expert in the subject matter. Furthermore, studies do not observe, even if qualitatively, how much experience the participant has with artificial intelligence. Other studies observing AI-generated images, for example, have higher rates of misclassification (38.7\%), as fake images can also emerge to be a problem in the future.

In addition, AI generated faces are now ``indisguishable'' from human faces due to advancements in generative adversarial networks; this can lead to the weaponization of artificially generated media~\cite{nightingale2022ai}. A 124 adult study observed whether these faces can be accurately classified as ``human'' or ``computer-generated''. Apart from classifying the face, participants also rated their confidence on a 0-100 scale. It was found that participants who had the least accuracy in detection simultaneously overestimated their confidence at this task (the Dunning-Kruger Effect), resulting in a higher misclassification rate.

Although social media platforms and their usage across age groups have been analyzed, including Facebook, Twitter, Instagram, Pinterest and LinkedIn~\cite{perrin2015social}, and forensic analyses of platforms such as Discord have taken place~\cite{moffitt2021discord}, there is a lack of focus on artificial intelligence and its implications on users. Discord also allows users to set up their own bots that can perform automated actions including moderation; this is done through the Discord API~\cite{abhinand2022study}. With existing chatbots (such as Discord's most popular MEE6) now incorporating artificial intelligence, it is plausible that it will only continue to be more commonplace.

Generation Z is growing up in an era where technologies such as automation, artificial intelligence, and machine learning are exponentially advancing and rapidly transforming societies and industries~\cite{wang2019artificial}. In order to maximise the benefits of these technologies while curtailing potential dangers, especially in misinformation and impersonation, it is important to study if Generation Z can determine the authenticity of AI-generated media. Understanding the perceptive acuity of this ``tech-savvy'' generation in distinguishing between the two sources can shed light on the effectiveness of AI-generated text and its potential impact on user trust, engagement, and interaction dynamics~\cite{miller2023ai}.

\section{Methods}
\label{sec:methods}

\subsection{Participants}

\subsubsection{Recruitment}

1882 individuals were identified from a science, technology, engineering, and math (STEM) education Discord server (a term for a community of users) with 67\% of individuals hailing from the United States, and 78\% being a member of the community for a considerable amount of time (over one month), as seen in Table \ref{tab:stemserver_demographics}. The individuals in the server and study are primarily middle school, high school, and college students, within the Generation Z age range (See Table \ref{tab:demographicstable}). Most students in the server have a passion for STEM research—their shared purpose for being in the server.

An invitation was sent from July to August of 2023 to the users of this server, providing an opt-in opportunity to participate. Users were consented upon registering for the study, and they were subsequently provided the questions to answer on the survey.

\begin{table}[h]
\begin{tabular}{|l|l|l|l|}
\hline
                                     &                   & Number of Samples & Percentage \\ \hline
\multirow{4}{*}{Location}            & United States     & 1262              & 67\%       \\ \cline{2-4} 
                                     & Other             & 385               & 20\%       \\ \cline{2-4} 
                                     & India             & 163               & 9\%        \\ \cline{2-4} 
                                     & Canada            & 72                & 4\%        \\ \hline
\multirow{3}{*}{Membership Duration} & 6-12 Months       & 751               & 40\%       \\ \cline{2-4} 
                                     & 1-6 Months        & 708               & 38\%       \\ \cline{2-4} 
                                     & \textless 1 Month & 423               & 22\%       \\ \hline
\end{tabular}
\caption{Discord Community Demographics from server analytics API}
\label{tab:stemserver_demographics}
\end{table}

\subsubsection{Demographics}

335 participants took part in the study, and achieved a margin of error of 5.35\% with confidence level of 95\%.
The participants' demographic information was collected through a survey and included the following variables:
\begin{enumerate}
	\item Age: Participants were asked to provide their age, which allowed for identifying Generation Z aged individuals, roughly born between 1997 and 2012.
	\item Gender: Participants indicated their gender, allowing for an understanding of gender distribution within the sample.
	\item Education background: Participants were asked about their highest level of education completed, providing insights into the educational diversity of the sample, and its influence on results.
	\item Continent of residence: Participants specified the continent in which they currently reside (for the past 5 years), helping to identify any potential regional variations in AI perception.
	\item Familiarity with Discord: Participants self-reported their familiarity with Discord and the frequency of their usage, to observe potential correlations between usage of Discord and accuracy of detection of AI generated text. Users were stratified into the categories:
    \begin{itemize}
    \item Never used Discord.com
    \item Novice (uses Discord 1-2 days/week)
    \item Proficient user (uses Discord 2-4 days/week)
    \item Frequent user (uses Discord 5-7 days/week)

    \end{itemize}
	\item Familiarity with AI: Participants self-reported their familiarity with AI technology, enabling the categorization of participants based on their knowledge and exposure to artificial intelligence.
    The options users had were:
    \begin{itemize}
        \item I have not heard of this before
        \item I have heard of this before, but have not used it
        \item I have used this before, but lack an understanding of how it works
        \item I have used this, and roughly understand how it works
        \item I have used this, and thoroughly understand how it works
    \end{itemize}
\end{enumerate}

The resultant distributions are depicted in Table \ref{tab:demographicstable}.

\begin{table}[H]
\centering
\begin{tabular}{|l|l|r|r|}
\hline
\textbf{} & \textbf{} & \textbf{No.} & \textbf{Percentage} \\ \hline
Age & 13-15 & 147 & 44.01\% \\ \cline{2-4} 
& 16-18 & 176 & 52.69\% \\ \cline{2-4} 
& 19-23 & 11 & 3.29\% \\ \hline
Gender & Female & 114 & 34.13\% \\ \cline{2-4} 
& Male & 211 & 63.17\% \\ \hline
Education & In college/undergraduate & 32 & 9.58\% \\ \cline{2-4} 
& In graduate school & 4 & 1.20\% \\ \cline{2-4} 
& In high school & 274 & 82.04\% \\ \cline{2-4} 
& In middle school & 21 & 6.29\% \\ \hline
Continent & Africa & 4 & 1.20\% \\ \cline{2-4} 
& Asia & 62 & 18.56\% \\ \cline{2-4} 
& Europe & 20 & 5.99\% \\ \cline{2-4} 
& North America & 249 & 74.55\% \\ \cline{2-4} 
& Oceania & 2 & 0.60\% \\ \cline{2-4} 
& South America & 1 & 0.30\% \\ \hline
Discord Experience & Frequent user & 180 & 53.89\% \\ \cline{2-4} 
& Never used Discord.com & 12 & 3.59\% \\ \cline{2-4} 
& Novice & 47 & 14.07\% \\ \cline{2-4} 
& Proficient user & 99 & 29.64\% \\ \hline
Experience with AI & Heard of, but not used & 12 & 3.59\% \\ \cline{2-4} 
& Not heard of & 2 & 0.60\% \\ \cline{2-4} 
& Used, but lack an understanding & 40 & 11.98\% \\ \cline{2-4} 
& Used and roughly understand & 169 & 50.60\% \\ \cline{2-4} 
& Used and thoroughly understand & 115 & 34.43\% \\ \hline
\end{tabular}
\caption{Demographics table based on participant survey}
\label{tab:demographicstable}
\end{table}

\subsubsection{Informed Consent}
Before participating in the study, participants were presented with an informed consent form outlining the purpose of the research, the procedures involved, and the confidentiality of their responses. They were informed that their participation was voluntary and that they could withdraw at any time without any consequences. Only those who provided their informed consent were permitted to take part in the study.

\subsection{Experimental Task and Discord Messages}
The experiment assessed  the participants’ perceptual acuity in identifying AI-generated content within the context of Discord messages. To create a diverse set of AI-generated Discord messages, the `gpt-3.5-turbo' language model was utilized via OpenAI API. Each message was generated using the same one-shot  prompt, where \texttt{\${hobby}} refers to one of 25 hobbies:
\begin{verbatim}
You are someone with a ${hobby} hobby.
Please describe your hobby ${hobby} in 3 sentences.
\end{verbatim}
A total of 25 different hobbies were incorporated into the prompt to ensure a comprehensive evaluation of participants’ ability to distinguish between AI-generated and human-authored text under various themes. For each of the 25 hobbies, an AI-generated message and human-authored example was created, resulting in a total of 50 messages. The human-authored messages involved a consistent use of prose and ordinary grammatical structures, characterized by coherence, clarity, and effective use of language. Different hobbies involve distinct terminologies and contexts. Utilizing a wide range of hobbies allows the study to assess participants' ability to recognize nuances specific to each hobby and determine whether they can distinguish AI-generated text from human-written content based on these subtleties. In addition, by focusing on hobbies, which are often discussed and shared on Discord, this reflects real-world conversation scenarios where users encounter AI-generated and human-authored content.

\subsection{Webhook and Image Capture}
\begin{figure}[h]
  \centering
  \includegraphics[width=\textwidth]{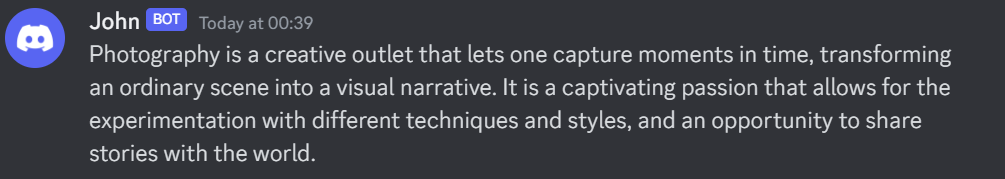}
  \caption{A sample image that was shown to  participants}
  \label{fig:webhookimg}
\end{figure}

Each AI-generated and human-authored Discord message was then posted through a webhook on an isolated Discord server. This allowed the messages to appear as though they were sent in a real Discord chat. To ensure consistency and avoid any potential bias in the presentation of messages, the same username and Discord-default profile image was used. The resultant examples were randomized and incorporated into the survey.

\subsection{Typeform Survey}
A survey was programmed to present participants with a ran-domized grouping of Discord messages for evaluation via Typeform. Each participant was shown 10 Discord scenarios from a pool of 50, with a random amount of human-authored/AI-generated messages. The messages were randomly selected and presented in a randomized order for each participant. Participants were tasked with classifying  each scenario presented as either ``AI-generated'' or ``Human-authored,'' via two options ``Human'' and ``AI.''

\section{Results}

\subsection{Effect of Experience with AI}

One of the variables studied was the self-reported experience individuals had with artificial  intelligence. Table \ref{tab:contingency_table} demonstrates significant deviations in the accuracy for artificial intelligence and humans. 

\begin{table}[H]
\centering
\begin{tabular}{lccc}
\hline
\textbf{Experience} & \textbf{AI Correct} & \textbf{Human Correct} \\
\hline
Used \& Thoroughly Understand & 49\% & 39\% \\
Used \& Roughly Understand & 47\% & 48\% \\
Used \& Lack Understanding & 50\% & 45\% \\
Heard of \& Not Used & 49\% & 42\% \\
Not Heard of & 45\% & 67\% \\
\hline
\end{tabular}
\caption{Contingency table (raw percentage) across user experience with AI}
\label{tab:contingency_table}
\end{table}

\begin{figure}[h]
    \centering
    \begin{subfigure}{0.3\textwidth}
        \includegraphics[width=\linewidth]{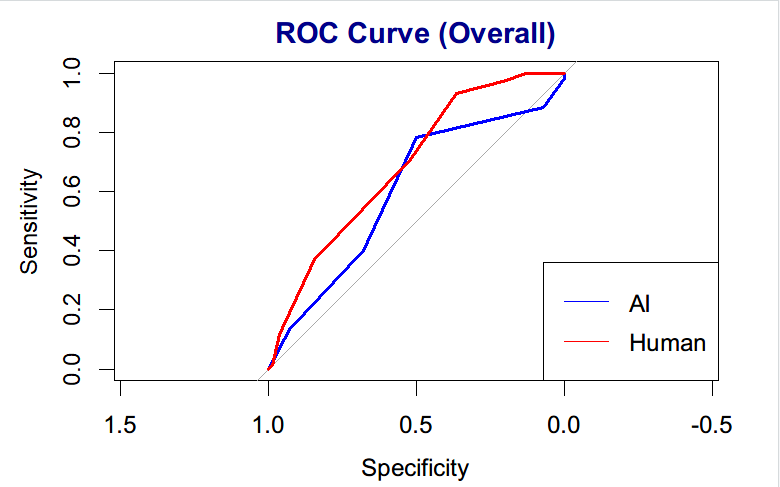}
        \caption{Overall Performance}
    \end{subfigure}
    \begin{subfigure}{0.3\textwidth}
        \includegraphics[width=\linewidth]{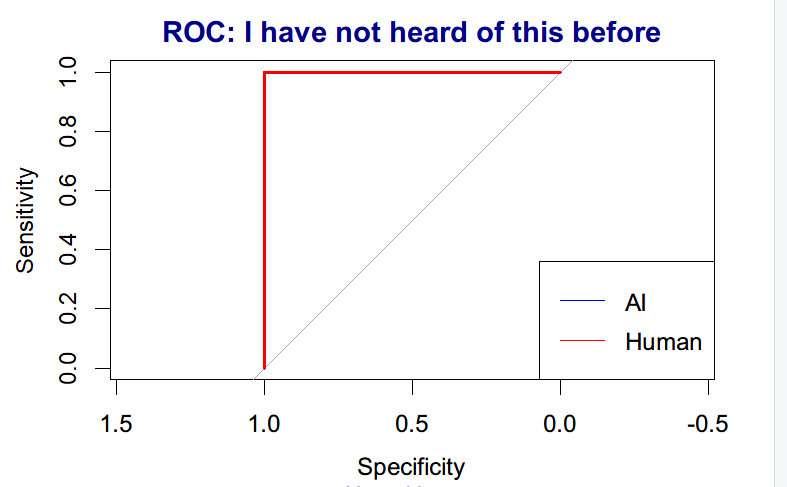}
        \caption{Not Heard Of AI}
    \end{subfigure}
    \begin{subfigure}{0.3\textwidth}
        \includegraphics[width=\linewidth]{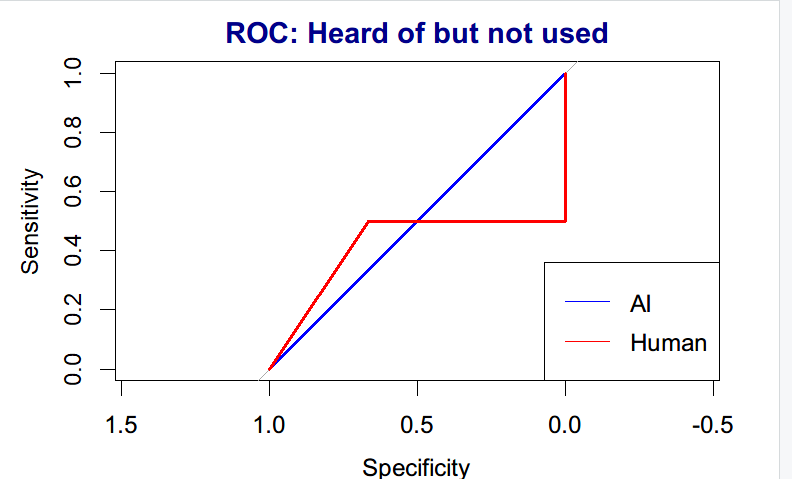}
        \caption{Heard Of \& Not Used AI}
    \end{subfigure}

    \medskip

    \begin{subfigure}{0.3\textwidth}
        \includegraphics[width=\linewidth]{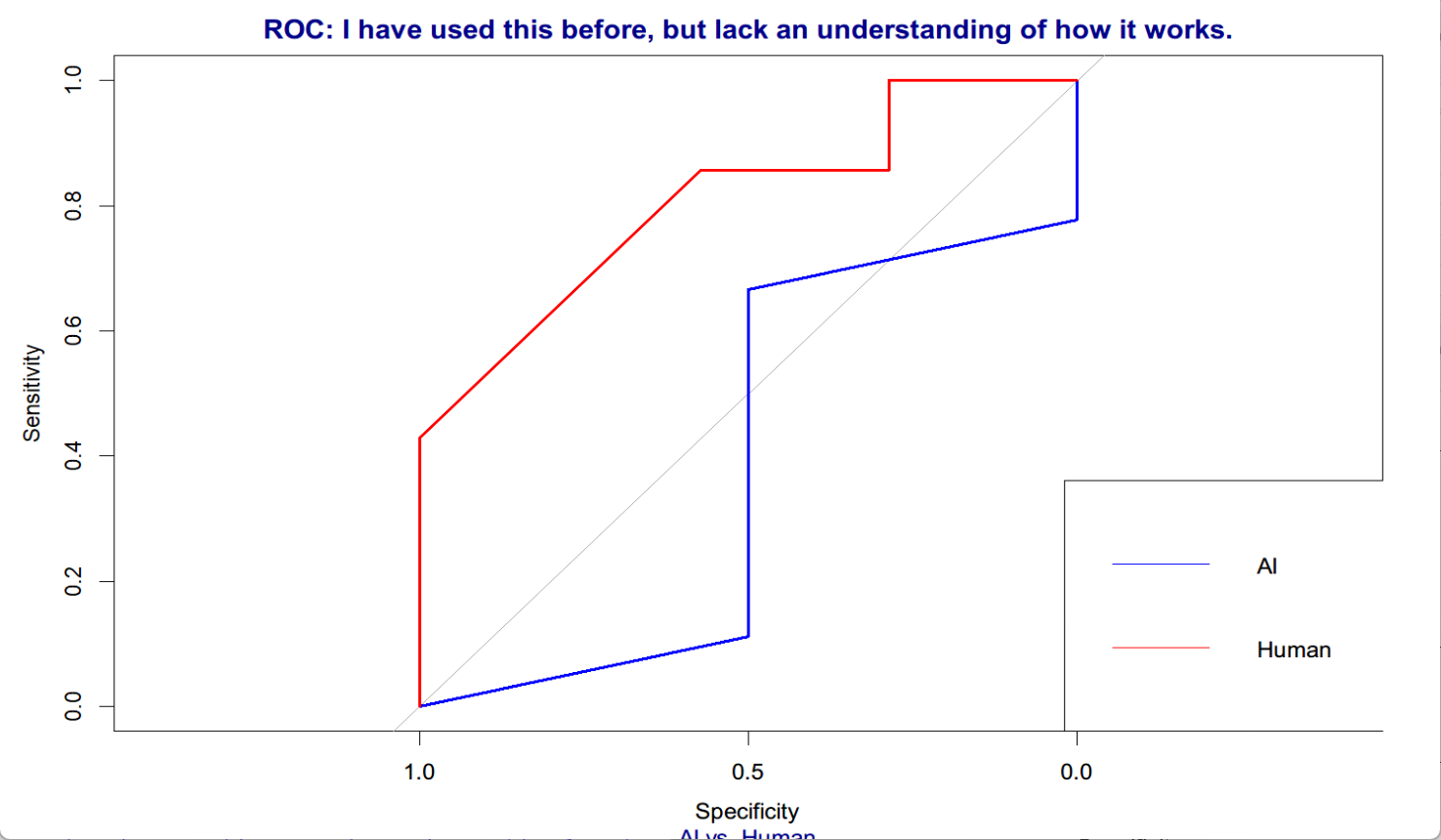}
        \caption{Lack Understanding of AI}
    \end{subfigure}
    \begin{subfigure}{0.3\textwidth}
        \includegraphics[width=\linewidth]{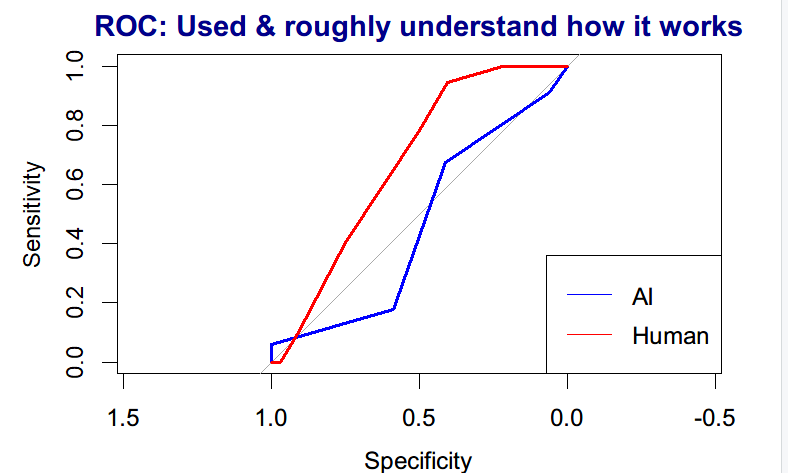}
        \caption{Roughly Understand AI}
    \end{subfigure}
    \begin{subfigure}{0.3\textwidth}
        \includegraphics[width=\linewidth]{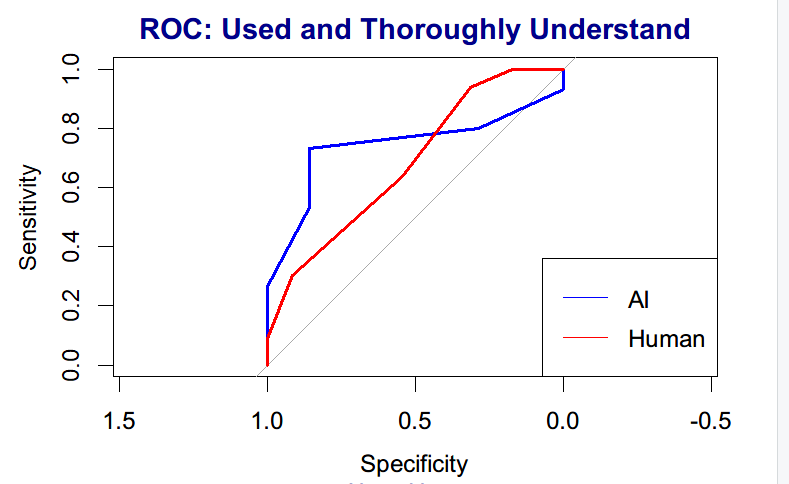}
        \caption{Thoroughly Understand AI}
    \end{subfigure}

    \caption{ROC curves for varying experience levels with AI}
    \label{fig:roc_curves}
\end{figure}

In the context of evaluating the classification of 335 samples encompassing both AI-generated and human-made text against random classification, the Receiver Operating Characteristic (ROC) curve provides a comprehensive visualization of the samples' ability, where a sample with a curve that ascends sharply towards the upper left corner is indicative of higher sensitivity and lower false positive rates. The area under the ROC curve (AUC) serves as a quantitative metric, with a higher AUC value signifying more accurate discriminatory ability. By comparing against random classification, the extent to which Generation Z aged individuals can discriminate text can be observed. As seen in Figure \ref{fig:roc_curves} (a), both curves are above the random classification mark. It can be seen for those with less experience and knowledge of artificial intelligence---(d) and (e)---classification of human-made text is more accurate than that of artificially generated text which falls below the random classification threshold.

Notably, for AI-generated texts, individuals with past AI usage but lacking understanding exhibited the highest accuracy (50\%), closely followed by those with a thorough understanding (49\%), and individuals completely unfamiliar with AI (49\%). Although accuracy in AI-generated text had little deviation, human-generated text classification demonstrated that those who had never heard of AI achieved the highest accuracy (67\%), while those with a thorough understanding exhibited the lowest accuracy (39\%). 

A paired t-test was conducted in order to observe a statistically significant difference between the AI experience and number of correct AI classifications. The t-statistic is 13.796 with 337 degrees of freedom, resulting in a p-value of $< 2.2 \times 10^{-16}$. This provides strong evidence to reject the null hypothesis that the true mean difference is equal to 0. The paired t-test between number of correct identifications of human-made text and experience with AI reveals a highly significant difference between AI experience and the number of correct answers. With a t-statistic of 13.796 and 337 degrees of freedom, the associated p-value is $< 2.2 \times 10^{-16}$, and we reject the null hypothesis.


\subsection{Demographic Analysis}

\begin{figure}[h]
  \centering
  \begin{subfigure}[b]{0.45\textwidth}
    \includegraphics[width=\textwidth]{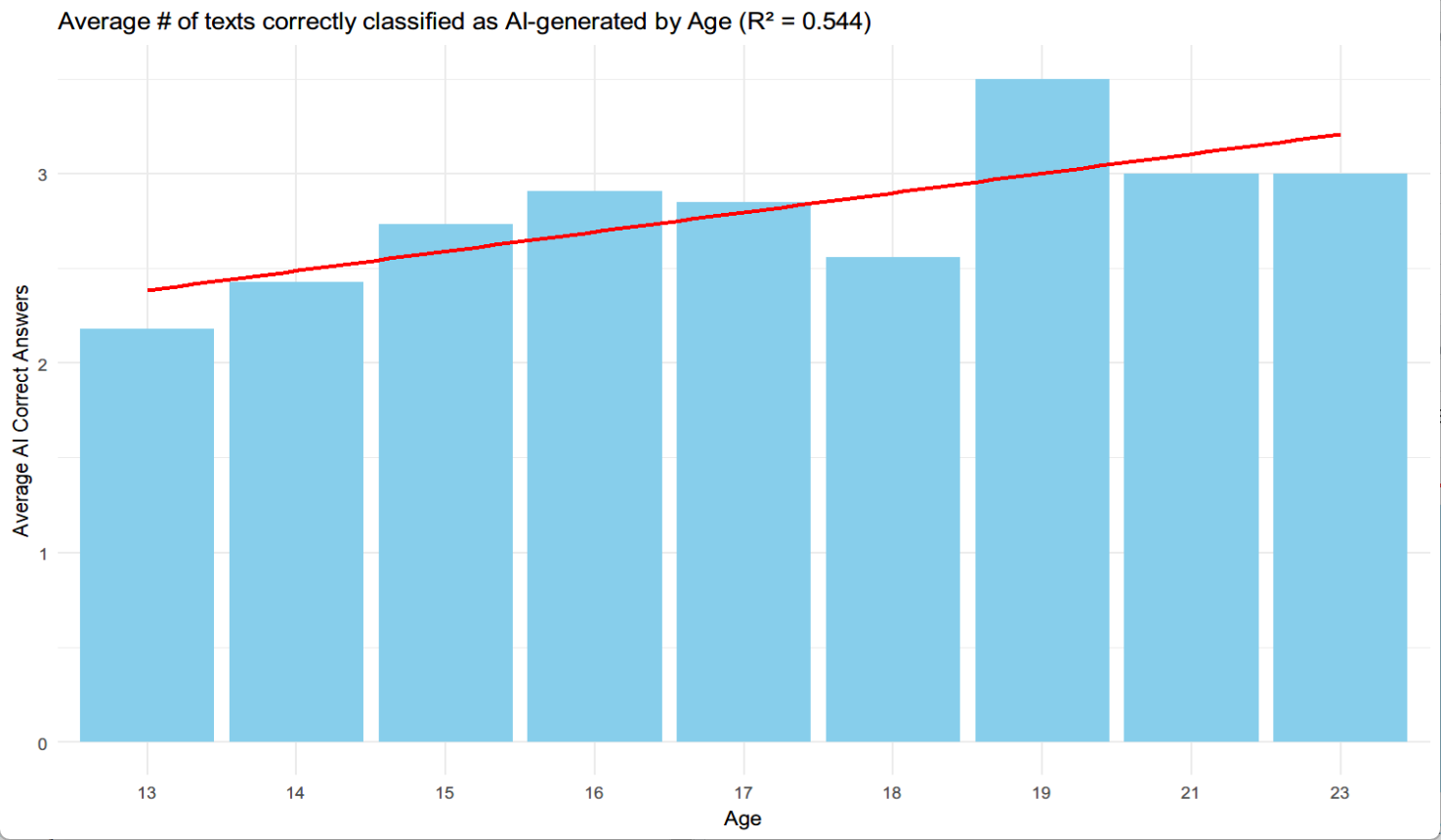}
    \caption{AI-generated text}
    \label{fig:avgaicorrect}
  \end{subfigure}
  \hfill
  \begin{subfigure}[b]{0.45\textwidth}
    \includegraphics[width=\textwidth]{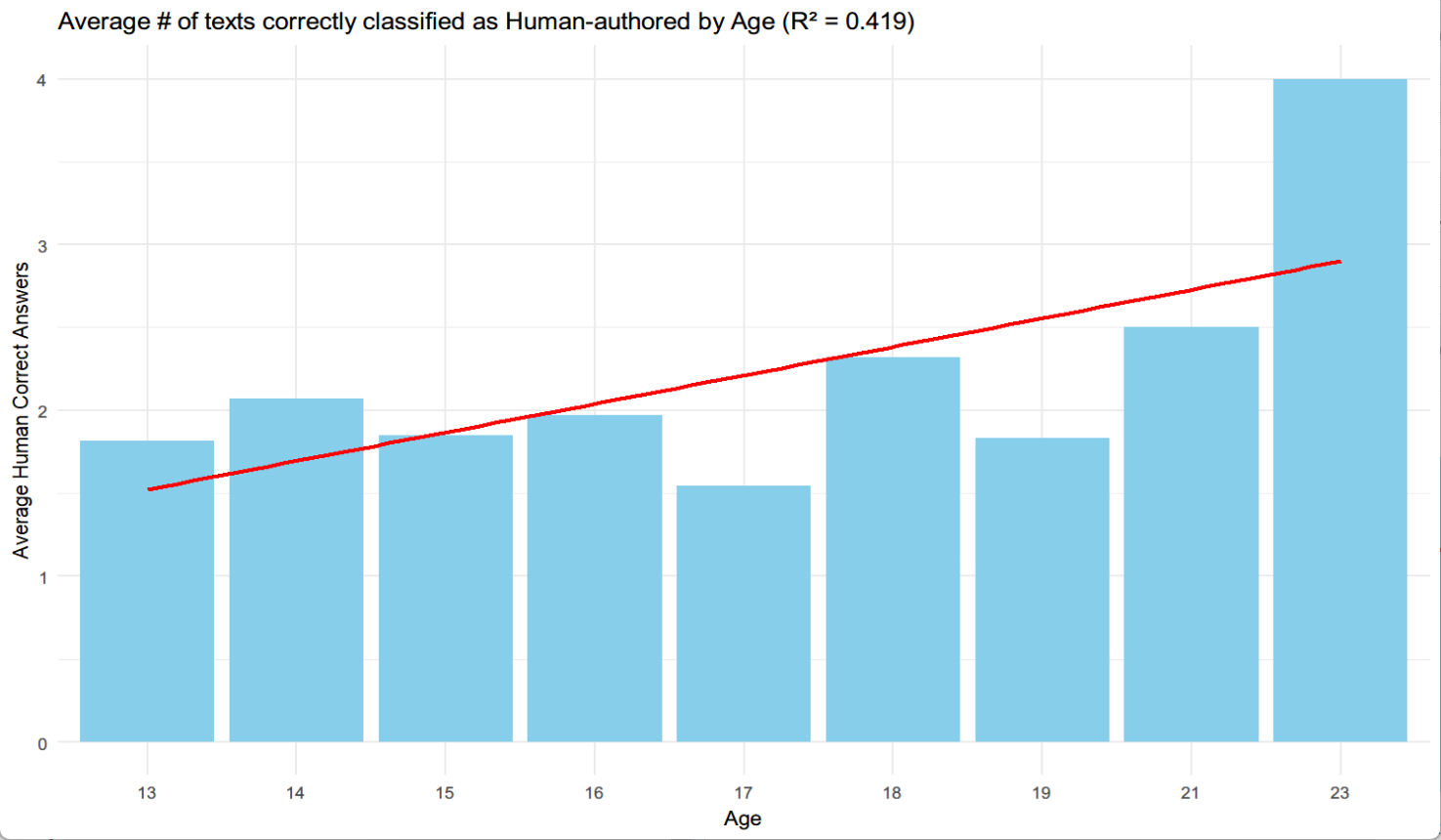}
    \caption{Human-created text}
    \label{fig:avghumancorrect}
  \end{subfigure}
  \caption{Comparison of correct identifications by age for AI-generated and human-created text}
  \label{fig:comparison}
\end{figure}

In the conducted study, participants were queried regarding their age, with a focus on individuals aged 13 and above. The demographic data as depicted in Table \ref{tab:demographicstable} revealed a noteworthy correlation, where the average number of correct classifications as human and AI increased with age. Figure \ref{fig:comparison} illustrates the abilities of participants to correctly classify text depending on their age. \\


Figure \ref{fig:avgaicorrect} illustrates the correlation between the average number of correct classifications of AI-generated text as AI and age, showing a bar chart accompanied by a linear regression model with the equation $\text{Average AI Correct} = 2.2815 + 0.1028(age)$. Figure \ref{fig:avghumancorrect} shows a steeper line, with the equation $\text{Average Human Correct} = 1.3516 + 0.1721(age)$ that indicates a greater range in correct answers across the age groups. However, when an analysis of variance (ANOVA) was conducted for the same data, it yielded a p-value of $0.17$ which was statistically insignificant.

Apart from age, the education level of the sample was also studied, in order to observe the potential effect on their accuracy in human or AI text classification. Individuals in middle school, high school, and those in college had accuracies of 41.90\%, 55.40\%, and 60.63\% respectively. Again, a similar tendency as seen in age is  observed, where higher education levels correspond to a higher accuracy when identifying artificially generated text. With regard to identifying human-made text, the average accuracy for those in middle school, high school, and college was 35.24\%, 37.96\% and 39.38\% respectively. There is a significant reduction in accuracy in identifying human-authored text, as well as lower deviation between education groups. A result of ANOVA, however, yielded a p-value of $0.204$.

Gender did not have a statistically significant impact on results (p = $0.218$). Individuals who identified as ”Male” and “Female” had accuracies of 55.83\% and 52.11\%, respectively. Individuals who identified as non-binary (n=3) did not have statistically significant conclusions.


\begin{figure}[h]
  \centering
  \includegraphics[width=0.48\textwidth]{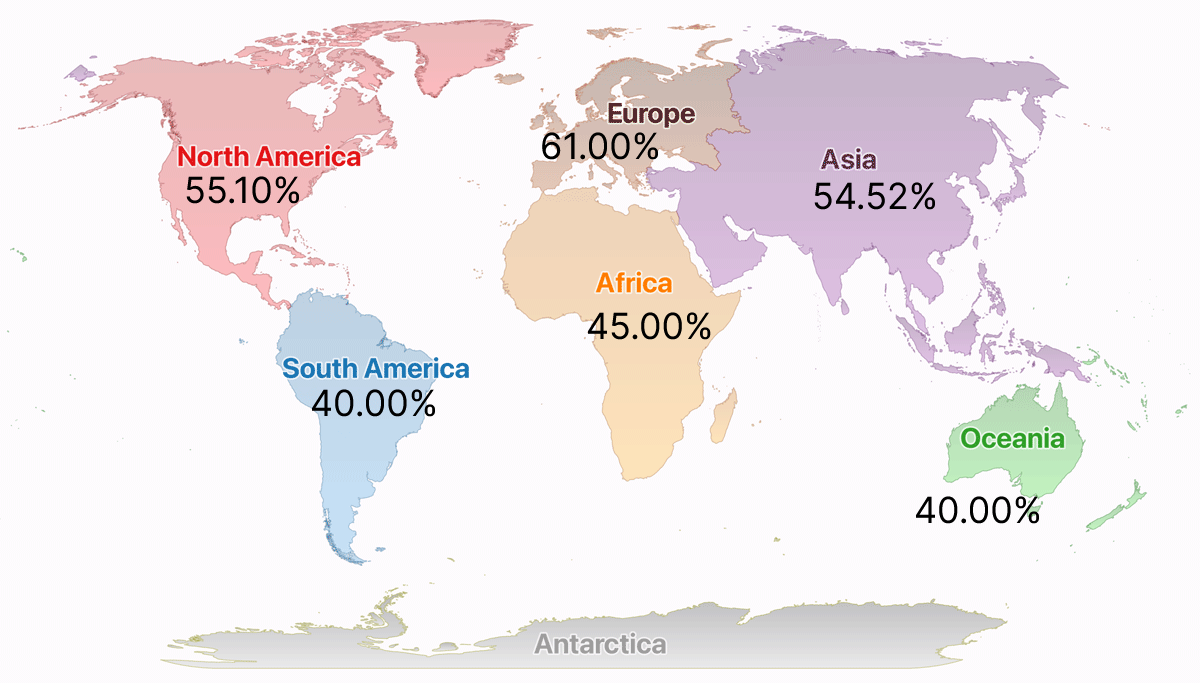}
  \hfill
  \includegraphics[width=0.48\textwidth]{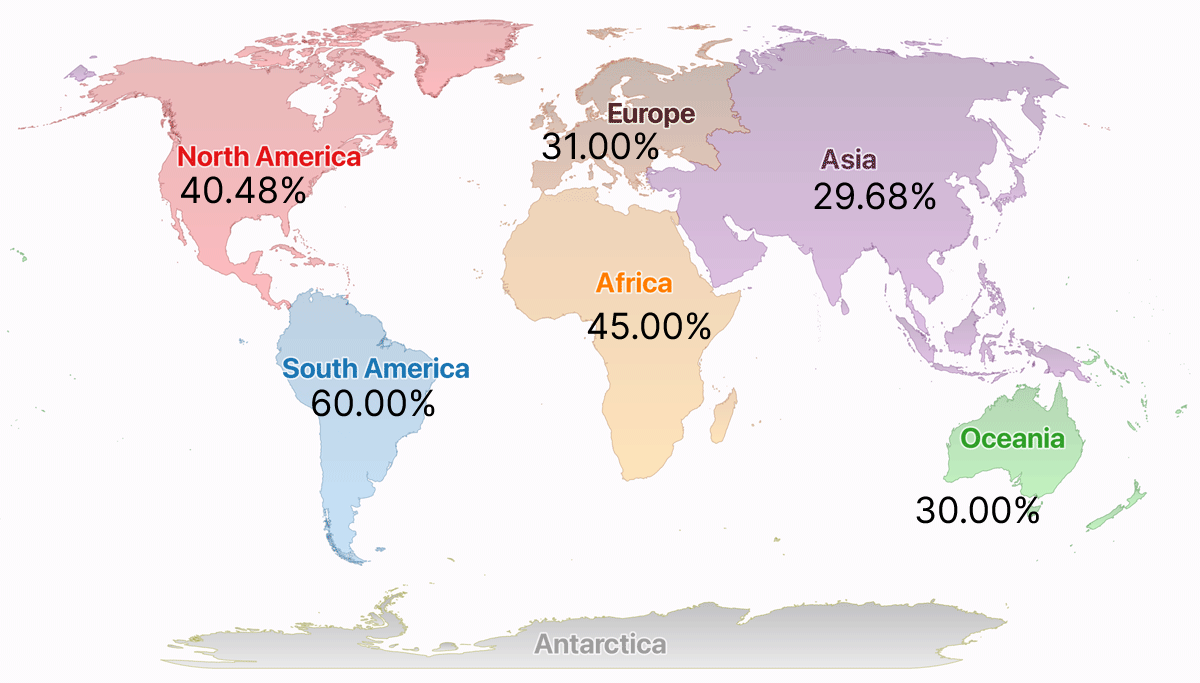}
  \caption{Comparison of percentage of correct classifications of AI-generated text (left) and human-written text (right) by continent}
  \label{fig:accuracy_comparison}
\end{figure}

\section{Limitations and Conclusion}

Given the significant association between experience with AI and classification of answers, revealed through the above analyses (p $< 2.2 \times 10^{-16}$), we conclude that overconfidence or bias stemming from perceived expertise in AI could exert an influence on individuals' judgments when classifying texts as human or AI generated. This is congrous with prior literature observing the classification of AI-generated faces~\cite{miller2023ai}. The Dunning-Kruger effect, a cognitive bias wherein individuals with lower ability at a task tend to overestimate their competence, is exemplified in the observed results~\cite{kruger1999unskilled}. Despite supposedly having a higher level of AI understanding, individuals may fall victim to this bias, erroneously presuming a more accurate discernment of human and AI-generated content than objectively warranted. The results also suggest anchoring bias, a specific cognitive bias where individuals rely too heavily on the first piece of information encountered (in this case, their expectations of AI-generated text), contributing to an increased likelihood of false positives. 

While the utilization of self-reporting in assessing participants' familiarity with AI technology served as a practical means to categorize individuals based on their knowledge and exposure, it is essential to acknowledge the inherent limitations and potential biases associated with this method. Self-reporting relies on participants' subjective perceptions of their own familiarity, introducing the possibility of bias and overestimation of their own skills. Moreover, individuals may have varying interpretations of what constitutes "familiarity" with AI, leading to subjective and potentially inconsistent categorizations. An extension to the research would involve an objective means of assessing their experience with artificial intelligence.

The findings of this study carry substantial implications for diverse domains, notably impacting AI development, education, and user experience. The discerned correlation between age and accuracy in distinguishing AI-generated text implies a developmental trajectory in individuals' perceptual acuity. It suggests that, on average, younger individuals may exhibit a higher susceptibility to misinterpreting AI-generated content as human-authored, especially in the context of prevalent social media platforms. This has inherent risks, as it means minors can be exposed to deceptive or manipulative content. The lack of a discernible correlation between gender, continent, educational background and accuracy in discriminating Human or AI-generated text underscores the universality of the challenge faced by Generation Z individuals. In the context of Discord as a social media platform, these considerations become integral to fostering a safe environment.

This study functions as a contemporary Turing test for social media platforms, probing the boundaries of artificial intelligence capabilities in emulating human behavior and interactions. As AI continues to advance, there is a foreseeable trajectory toward achieving Artificial General Intelligence (AGI), a state where machines can perform any intellectual task that a human being can. The envisioned future scenario implies a seamless integration of AI into social media, where the distinction between AI-driven entities and human users becomes increasingly imperceptible. The evolution toward AGI prompts contemplation on the potential convergence of AI and human behavioral patterns within the digital realm, thereby challenging traditional notions of discerning AI from genuine human engagement.



 \bibliographystyle{elsarticle-num} 
 \bibliography{cas-refs}

\begin{thebibliography}{10}
\expandafter\ifx\csname url\endcsname\relax
  \def\url#1{\texttt{#1}}\fi
\expandafter\ifx\csname urlprefix\endcsname\relax\def\urlprefix{URL }\fi
\expandafter\ifx\csname href\endcsname\relax
  \def\href#1#2{#2} \def\path#1{#1}\fi

\bibitem{Makridakis_2017}
S.~Makridakis, The forthcoming artificial intelligence (ai) revolution: Its
  impact on society and firms, Futures 90 (2017) 46–60.
\newblock \href {https://doi.org/10.1016/j.futures.2017.03.006}
  {\path{doi:10.1016/j.futures.2017.03.006}}.

\bibitem{Gupta_Varol_Zhou_2023}
K.~Gupta, C.~Varol, B.~Zhou, Digital forensic analysis of discord on google
  chrome, Forensic Science International: Digital Investigation 44 (2023)
  301479.
\newblock \href {https://doi.org/10.1016/j.fsidi.2022.301479}
  {\path{doi:10.1016/j.fsidi.2022.301479}}.

\bibitem{bian2023drop}
N.~Bian, P.~Liu, X.~Han, H.~Lin, Y.~Lu, B.~He, L.~Sun, A drop of ink makes a
  million think: The spread of false information in large language models
  (2023).
\newblock \href {http://arxiv.org/abs/2305.04812} {\path{arXiv:2305.04812}}.

\bibitem{dimock2019defining}
M.~Dimock, Defining generations: Where millennials end and generation z begins,
  Pew Research Center 17~(1) (2019) 1--7.

\bibitem{Vinichenko-viewsofgenz}
V.~et~al., Threats and risks from the digitalization of society and artificial
  intelligence: Views of generation z students, International Journal of
  ADVANCED AND APPLIED SCIENCES 8~(10) (2021) 108–115.
\newblock \href {https://doi.org/10.21833/ijaas.2021.10.012}
  {\path{doi:10.21833/ijaas.2021.10.012}}.

\bibitem{whyte2020deepfake}
C.~Whyte, Deepfake news: Ai-enabled disinformation as a multi-level public
  policy challenge, Journal of cyber policy 5~(2) (2020) 199--217.

\bibitem{Ceci_2023}
L.~Ceci,
  \href{https://www.statista.com/statistics/1367922/discord-registered-users-worldwide/}{Discord
  global mau 2023} (Jul 2023).
\newline\urlprefix\url{https://www.statista.com/statistics/1367922/discord-registered-users-worldwide/}

\bibitem{9575079}
M.~L. Arifianto, I.~F. Izzudin, From gaming to learning: Assessing the
  gamification of discord in the realm of education, in: 2021 7th International
  Conference on Education and Technology (ICET), 2021, pp. 95--99.
\newblock \href {https://doi.org/10.1109/ICET53279.2021.9575079}
  {\path{doi:10.1109/ICET53279.2021.9575079}}.

\bibitem{moffitt2021discord}
K.~Moffitt, U.~Karabiyik, S.~Hutchinson, Y.~H. Yoon, Discord forensics: The
  logs keep growing, in: 2021 ieee 11th annual computing and communication
  workshop and conference (ccwc), IEEE, 2021, pp. 0993--0999.

\bibitem{abhinand2022study}
G.~Abhinand, R.~Balasubramanian, Study on the development and implementation of
  ubiquitous bots for the discord interface (2022).

\bibitem{dale_2021}
R.~Dale, Gpt-3: What’s it good for?, Natural Language Engineering 27~(1)
  (2021) 113–118.
\newblock \href {https://doi.org/10.1017/S1351324920000601}
  {\path{doi:10.1017/S1351324920000601}}.

\bibitem{clyde}
\href{https://support.discord.com/hc/en-us/articles/13066317497239-Clyde-Discord-s-AI-Chatbot}{Clyde:
  Discord's ai chatbot}.
\newline\urlprefix\url{https://support.discord.com/hc/en-us/articles/13066317497239-Clyde-Discord-s-AI-Chatbot}

\bibitem{jain2023generative}
R.~Jain, A.~Jain, Generative ai in writing research papers: A new type of
  algorithmic bias and uncertainty in scholarly work (2023).
\newblock \href {http://arxiv.org/abs/2312.10057} {\path{arXiv:2312.10057}}.

\bibitem{ray2023chatgpt}
P.~P. Ray, Chatgpt: A comprehensive review on background, applications, key
  challenges, bias, ethics, limitations and future scope, Internet of Things
  and Cyber-Physical Systems (2023).

\bibitem{kocon2023chatgpt}
J.~Koco{\'n}, I.~Cichecki, O.~Kaszyca, M.~Kochanek, D.~Szyd{\l}o, J.~Baran,
  J.~Bielaniewicz, M.~Gruza, A.~Janz, K.~Kanclerz, et~al., Chatgpt: Jack of all
  trades, master of none, Information Fusion (2023) 101861.

\bibitem{kreps2022all}
S.~Kreps, R.~M. McCain, M.~Brundage, All the news that’s fit to fabricate:
  Ai-generated text as a tool of media misinformation, Journal of experimental
  political science 9~(1) (2022) 104--117.

\bibitem{denny2023can}
P.~Denny, H.~Khosravi, A.~Hellas, J.~Leinonen, S.~Sarsa, Can we trust
  ai-generated educational content? comparative analysis of human and
  ai-generated learning resources, arXiv preprint arXiv:2306.10509 (2023).

\bibitem{pegoraro2023chatgpt}
A.~Pegoraro, K.~Kumari, H.~Fereidooni, A.-R. Sadeghi, To chatgpt, or not to
  chatgpt: That is the question! (2023).
\newblock \href {http://arxiv.org/abs/2304.01487} {\path{arXiv:2304.01487}}.

\bibitem{chaka2023detecting}
C.~Chaka, Detecting ai content in responses generated by chatgpt, youchat, and
  chatsonic: The case of five ai content detection tools, Journal of Applied
  Learning and Teaching 6~(2) (2023).

\bibitem{openaiclassifier}
\href{https://platform.openai.com/ai-text-classifier}{{New AI classifier for
  indicating AI-written text}}, [Online; accessed 18. Dec. 2023] (Dec. 2023).
\newline\urlprefix\url{https://platform.openai.com/ai-text-classifier}

\bibitem{sadasivan2023aigenerated}
V.~S. Sadasivan, A.~Kumar, S.~Balasubramanian, W.~Wang, S.~Feizi, Can
  ai-generated text be reliably detected? (2023).
\newblock \href {http://arxiv.org/abs/2303.11156} {\path{arXiv:2303.11156}}.

\bibitem{gonzalez2022allying}
M.~F. Gonzalez, W.~Liu, L.~Shirase, D.~L. Tomczak, C.~E. Lobbe, R.~Justenhoven,
  N.~R. Martin, Allying with ai? reactions toward human-based, ai/ml-based, and
  augmented hiring processes, Computers in Human Behavior 130 (2022) 107179.

\bibitem{10.1145/3290605.3300469}
M.~Jakesch, M.~French, X.~Ma, J.~T. Hancock, M.~Naaman,
  \href{https://doi.org/10.1145/3290605.3300469}{Ai-mediated communication: How
  the perception that profile text was written by ai affects trustworthiness},
  in: Proceedings of the 2019 CHI Conference on Human Factors in Computing
  Systems, CHI '19, Association for Computing Machinery, New York, NY, USA,
  2019, p. 1–13.
\newblock \href {https://doi.org/10.1145/3290605.3300469}
  {\path{doi:10.1145/3290605.3300469}}.
\newline\urlprefix\url{https://doi.org/10.1145/3290605.3300469}

\bibitem{10.1145/3334480.3382842}
T.~Bruzzese, I.~Gao, G.~Dietz, C.~Ding, A.~Romanos,
  \href{https://doi.org/10.1145/3334480.3382842}{Effect of confidence
  indicators on trust in ai-generated profiles}, in: Extended Abstracts of the
  2020 CHI Conference on Human Factors in Computing Systems, CHI EA '20,
  Association for Computing Machinery, New York, NY, USA, 2020, p. 1–8.
\newblock \href {https://doi.org/10.1145/3334480.3382842}
  {\path{doi:10.1145/3334480.3382842}}.
\newline\urlprefix\url{https://doi.org/10.1145/3334480.3382842}

\bibitem{cheng2022human}
X.~Cheng, X.~Zhang, J.~Cohen, J.~Mou, Human vs. ai: Understanding the impact of
  anthropomorphism on consumer response to chatbots from the perspective of
  trust and relationship norms, Information Processing \& Management 59~(3)
  (2022) 102940.

\bibitem{folstad2018makes}
A.~F{\o}lstad, C.~B. Nordheim, C.~A. Bj{\o}rkli, What makes users trust a
  chatbot for customer service? an exploratory interview study, in: Internet
  Science: 5th International Conference, INSCI 2018, St. Petersburg, Russia,
  October 24--26, 2018, Proceedings 5, Springer, 2018, pp. 194--208.

\bibitem{kobis2021artificial}
N.~K{\"o}bis, L.~D. Mossink, Artificial intelligence versus maya angelou:
  Experimental evidence that people cannot differentiate ai-generated from
  human-written poetry, Computers in human behavior 114 (2021) 106553.

\bibitem{gao2023comparing}
C.~A. Gao, F.~M. Howard, N.~S. Markov, E.~C. Dyer, S.~Ramesh, Y.~Luo, A.~T.
  Pearson, Comparing scientific abstracts generated by chatgpt to real
  abstracts with detectors and blinded human reviewers, NPJ Digital Medicine
  6~(1) (2023) 75.

\bibitem{nightingale2022ai}
S.~J. Nightingale, H.~Farid, Ai-synthesized faces are indistinguishable from
  real faces and more trustworthy, Proceedings of the National Academy of
  Sciences 119~(8) (2022) e2120481119.

\bibitem{perrin2015social}
A.~Perrin, Social media usage, Pew research center 125 (2015) 52--68.

\bibitem{wang2019artificial}
W.~Wang, K.~Siau, Artificial intelligence, machine learning, automation,
  robotics, future of work and future of humanity: A review and research
  agenda, Journal of Database Management (JDM) 30~(1) (2019) 61--79.

\bibitem{miller2023ai}
E.~J. Miller, B.~A. Steward, Z.~Witkower, C.~A. Sutherland, E.~G. Krumhuber,
  A.~Dawel, Ai hyperrealism: Why ai faces are perceived as more real than human
  ones, Psychological Science (2023) 09567976231207095.

\bibitem{kruger1999unskilled}
J.~Kruger, D.~Dunning, Unskilled and unaware of it: how difficulties in
  recognizing one's own incompetence lead to inflated self-assessments.,
  Journal of personality and social psychology 77~(6) (1999) 1121.

\end{thebibliography}





\end{document}